\documentclass[fleqn,usenatbib]{mnras}

\usepackage{newtxtext,newtxmath}
\usepackage[T1]{fontenc}
\usepackage{ae,aecompl}

\usepackage{graphicx}	% Including figure files
\usepackage{amsmath}	% Advanced maths commands
\usepackage{amssymb}	% Extra maths symbols

\newcommand{\acb}{$\alpha$ Cen B}
\newcommand{\logr}{$\log{R'_{HK}}$}
\newcommand{\angstrom}{\mbox{\normalfont\AA}}

\def\ms{{\rm m~s$^{-1}$}}

\title[Activity indicators with HARPS-N Solar Observations]{The spectral impact of magnetic activity on disk-integrated HARPS-N solar observations: exploring new activity indicators.}
\author[A. P. G. Thompson et al.]{
A. P. G. Thompson$^{1}$\thanks{E-mail:athompson1501@qub.ac.uk},
C. A. Watson$^{1}$,
R. D. Haywood$^{2\dagger}$,
J. C. Costes$^{1}$,
\newauthor
E. de Mooij$^{1,6}$,
A. Collier Cameron$^{3}$,
X. Dumusque$^{4}$,
D. F. Phillips$^{2}$,
\newauthor
S. H. Saar$^{2}$,
A. Mortier$^{5}$,
T. W. Milbourne$^{2}$,
S. Aigrain$^{9}$,
H. M. Cegla$^{4}$,
\newauthor
D. Charbonneau$^{2}$,
R. Cosentino$^{8}$,
A. Ghedina$^{8}$,
D. W. Latham$^{2}$,
\newauthor
M. L\'opez-Morales$^{1}$,
G. Micela$^{10}$,
E. Molinari$^{7}$,
E. Poretti$^{8}$,
A. Sozzetti$^{7}$,
\newauthor
S. Thompson$^{5}$,
and R. Walsworth$^{2}$
\\
$^{1}$Astrophysics Research Centre, School of Mathematics and Physics, Queen's University Belfast, BT7 1NN, Belfast, UK\\
$^{2}$Center for Astrophysics | Harvard \& Smithsonian, 60 Garden Street, Cambridge, MA 02138 USA\\
$^{3}$SUPA, School of Physics and Astronomy, University of St Andrews, North Haugh, St Andrews KY16 9SS, UK\\
$^{4}$Observatoire Astronomique de l'Universit\'{e} de Gen\'{e}ve, 51 Chemin des Maillettes, 1290 Sauverny, Suisse\\
$^{5}$Astrophysics Group, Cavendish Laboratory, University of Cambridge, J.J. Thomson Avenue, Cambridge CB3 0HE, UK\\
$^{6}$School of Physical Sciences and Centre for Astrophysics and Relativity, Dublin City University, Glasnevin, Dublin 9 , Ireland\\
$^{7}$INAF - Osservatorio Astrofisico di Torino, via Osservatorio 20, 10025 Pino Torinese, Italy\\
$^{8}$INAF-Fundacion Galileo Galilei, Rambla Jose Ana Fernandez Perez 7, E-38712 Brena Baja, Spain\\
$^{9}$Oxford Astrophysics, Denys Wilkinson Building, University of Oxford, OX1 3RH, Oxford, UK\\
$^{10}$INAF-Osservatorio Astronomico di Palermo, Piazza del Parlamento 1, I-90134 Palermo, Italy \\
$\dagger$NASA Sagan Fellow
}

\date{Accepted XXX. Received YYY; in original form ZZZ}

\pubyear{2019}

\begin{document}
\label{firstpage}
\pagerange{\pageref{firstpage}--\pageref{lastpage}}
\maketitle

\begin{abstract}
Stellar activity is the major roadblock on the path to finding true Earth-analogue planets with the Doppler technique. Thus, identifying new indicators that better trace magnetic activity (i.e. faculae and spots) is crucial to aid in disentangling these signals from that of a planet's Doppler wobble. In this work, we investigate activity related features as seen in disk-integrated spectra from the HARPS-N solar telescope. We divide high-activity spectral echelle orders by low-activity master templates (as defined using both \logr~and images from the Solar Dynamics Observatory, SDO), creating ``relative spectra''. With resolved images of the surface of the Sun (via SDO), the faculae and spot filling factors can be calculated, giving a measure of activity independent of, and in addition to, \logr. We find pseudo-emission (and pseudo-absorption) features in the relative spectra that are similar to those reported in our previous work on \acb. In \acb, the features are shown to correlate better to changes in faculae filling factor than spot filling factor. In this work we more confidently identify changes in faculae coverage of the visible hemisphere of the Sun as the source of features produced in the relative spectra. Finally, we produce trailed spectra to observe the RV component of the features, which show that the features move in a redward direction as one would expect when tracking active regions rotating on the surface of a star.
\end{abstract}

\begin{keywords}
techniques: radial velocities; Sun: activity; stars: activity; Sun: faculae, plages; 
planets and satellites: detection
\end{keywords}

\section{Introduction}

In \cite{Thompson2017}, and hereafter Paper~1, we investigated the impact of stellar activity from \acb~on HARPS spectra, by comparing observations taken during magnetically active and inactive phases. This work uncovered a forest of ``pseudo-emission features'': increases in the intensity of the relative spectra constructed by dividing `high activity' spectra (as defined by \logr) by a low activity template spectrum. These features varied in strength and velocity on timescales of the stellar rotation period, and also correlated well with the passage of a region of activity as seen in \logr. This work has since been verified by \cite{Wise2018} and \cite{Lisogorskyi2019}. In addition, in Paper~1, we found morphological differences in the pseudo-emission features that pointed to the underlying physics driving the line-profile changes. Simple modelling of the shapes and strengths of these pseudo-emission features strongly indicated that the spectral variability was likely being driven by faculae, rather than starspots.

This is potentially an important result in the field of radial-velocity (RV) measurements of extra-solar planets. While active stars are spot-dominated, quiet stars are known to be plage/faculae-dominated \citep{Radick1990, Hall2009, Lockwood2007, Shapiro2016}. Faculae and plage are the terms used to describe bright magnetic regions in stellar atmospheres, with faculae being used when describing these regions within the photosphere, and plage as its chromospheric counterpart. It is faculae that is believed to represent the fundamental barrier to the RV detection of, for example, an Earth analogue \citep{Saar1997, Saar2003, Meunier2010, Dumusque2014, Haywood2016}. Appearing as bright magnetic regions, faculae suppresses the underlying convection in the stellar atmosphere. Since convection consists of uprising, bright, hot material (granules) that cools and sinks into dark intergranular lanes, convective motion imprints a predominantly blue-shifted signal on spectral lines. Where faculae occurs, the associated enhanced magnetic field inhibits convection -- which suppresses the convective blueshift component relative to the surrounding photosphere. This leads to spurious RV signals as active regions both evolve and rotate across the stellar disk. The higher filling factor, more homogeneous distribution, and lower contrast of faculae compared to the much darker spots have made faculae difficult to track in both light-curves and spectra. As a result, techniques to remove their effects such as the FF' method \citep{Aigrain2012} assume that faculae and spots are spatially coincident, which then allows the light-curves to inform the RV noise solution (since the light-curve modulation of the spots can be monitored). In the case of the Sun, however, we know that this underlying assumption (that spots and faculae are spatially coincident) is not the case. For example, the longer lifetimes of facular regions relative to spots (see \citealt{Shapiro2020}, and referenes therein) mean they can persist for several solar rotations, whereas sunspots rarely complete more than 1 rotation. Thus, faculae are frequently unrelated to spots.

While there are a number of other sophisticated astrophysical noise removal techniques that perform well \citep[see][for a comprehensive breakdown of the most up-to-date techniques used within RV measurements]{Dumusque2017}, most do not make use of the wealth of information contained within a spectrum. The standard approach to mitigating RV noise from stellar activity is to decorrelate stellar noise from measured RVs using cross-correlation function (CCF) parameters. These include the CCF FWHM \citep[e.g.][]{Queloz2009, Hatzes2010} and bisector span \citep{Queloz2001} -- although recent work has demonstrated that using a Skewed-Normal distribution provides a more self-consistent result than measuring CCF parameters and bisector span separately \citep{Simola2019}. Other recent work has also tried to make better use of the high-resolution spectral information by instruments like HARPS. \cite{Dumusque2018}, for example, looked at the RV motion of individual lines rather than using a CCF. They were able to categorise spectral lines as either being sensitive or insensitive to changes in activity, allowing for either a reduction in the RV variation due to activity (i.e. better for planet detection) or to better isolate the RV variation due to activity to probe the surface of stars. Indeed, with our previous work in Paper~1, we see individual lines responding differently to the effects of activity, with some lines more affected by activity than others. This valuable activity information can become more entangled with planetary RV signals when using the averaging approach of the CCF.

The aim of this paper is to conduct a similar analysis to that of Paper~1 and determine whether individual spectral-line variability (driven by stellar activity) can also be identified in disk-integrated spectra of the Sun. This would also enable the underlying cause (sunspots or faculae) of such variability to be  determined by directly relating such variability to observable features on the solar surface. Determining the activity-sensitive lines in the Sun may also be of future use for uncovering an Earth-analogue planet. We obtained high-resolution, disk-integrated solar spectra using the HARPS-N solar telescope \citep{Phillips2016}, a small solar telescope that feeds light into the HARPS-N instrument on the TNG \citep{Cosentino2012}. In this study we perform a similar analysis to Paper~1 and follow the framework laid out in it: producing relative solar spectra to highlight any changes in the strength of spectral lines and correlate these with activity indicators like \logr, as well as spot- and faculae-filling factors determined via Solar Dynamics Observatory (SDO) images.

The paper is structured as follows. In Section \ref{sect:data}, we describe the HARPS-N Solar Telescope, as well as the data reduction process. This section also includes an investigation into, and removal of, a strong etaloning found in the data. In Section \ref{sect:features}, we perform a search of the data for features similar to those found in \acb~(Paper~1) and produce trailed spectral to better visualise the RV variations of the features. The results of this search are discussed and summarised in Section \ref{sect:discussion}.

\section{Data and Data Reduction}\label{sect:data}

Data for this study was obtained using the HARPS-N Solar Telescope. This is an instrument that takes full disk-integrated spectroscopic measurements of the Sun to better understand how surface inhomogeneities can affect precise RV measurements. Using its unique `Sun-as-a-star' observing capabilities, the RV signatures of stellar activity can be isolated from that of the (known) Doppler wobble shifts induced by the solar-system planets -- providing a clean trace of the astrophysical noise associated with solar activity. Only a brief description of the instrument is given here, see \cite{Phillips2016} for more detail on the telescope and \cite{Dumusque2015} for information on its science goals. The HARPS-N solar telescope is a small 3-inch telescope that takes light from the Sun and fibre feeds it to the HARPS-N instrument. Being the only setup to operate at the telescope during the day, the survey is run continuously and limited only by weather. A 5-minute exposure time is selected to reduce the effects of solar oscillations. Since seeing first light on $18^{\rm{th}}$ July 2015, the telescope has been continuously taking data, and at the time of writing there are over 60,000 high resolution, high signal-to-noise measurements of the Sun spanning over three years \citep[see ][ for the data of the first three years]{CollierCameron2019}.

\begin{figure}
    	\includegraphics[width=\columnwidth]{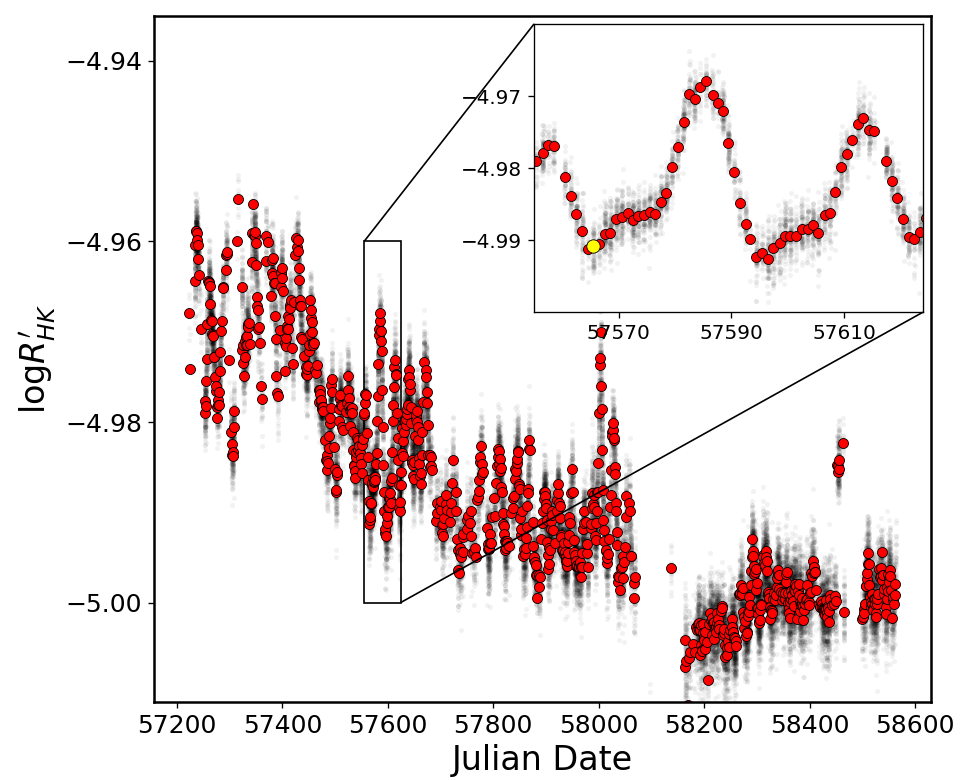}
    		\caption[\logr~measurements of the Sun]{HARPS-N solar telescope \logr~measurements of the Sun, where the variability seen is due to active regions rotating on and off the solar surface, with the red points representing the average value per day. Although a large amount of variation can be seen, all these values can be considered as being from a `quiet' star since, during this period, the Sun was on descent into solar minimum. The 2016 interval selected for analysis (MJD:$57555 - 57624$) is shown by the zoomed in region and chosen from the \logr~values. The day used for the low activity template is marked in yellow.}
    	\label{fig:Sun_logr}
    \end{figure}

For the analysis presented in this paper, a small subset of the data was selected based on the level of solar activity, and was processed in a similar way to Paper~1 to produce comparable results. The time interval selected contained 69 days worth of data from $16^{\rm{th}}$ June 2016 to $24^{\rm{th}}$ August 2016 (MJD:$57555 - 57624$). This 2016 interval was selected as this shows a clear periodic modulation of the \logr~over two solar rotation periods (highlighted in the embedded panel in Figure~\ref{fig:Sun_logr}) and exhibited one of the largest changes in activity over a single solar rotation period. For this analysis a low-activity template was created, and we selected data from $26^{\rm{th}}$ June 2016 for this purpose as it was one of the lowest activity days during the 2016 interval (highlighted as a yellow point in Figure~\ref{fig:Sun_logr}). The spectra from this day were compared to SDO/HMI images, which showed no spots on the surface and very little faculae, so represent a close approximation of the Sun having a truly immaculate photosphere.

We would like to note one difference between the analysis conducted in this work, and that conducted in Paper-1. In this work the low-activity template spectrum (against which all other spectra are compared in our analysis in order to monitor spectral-line changes) is taken from within the $16^{\rm{th}}$ June 2016 to $24^{\rm{th}}$ August 2016 period itself. This is different from our analysis presented in Paper-1, where the low-activity template spectrum used was from the minimum of \acb's activity cycle - which occurred some years before the `high-activity' period analysed. Thus, in the analysis of the solar-spectrum variations presented in this work, any variations observed are due to rotationally modulated activity (i.e. spots and faculae), and not due to activity-cycle changes. In addition, this will reduce the strength of the activity-related spectral changes we detect, since we have not maximised the activity-level differences between 'active' and 'inactive' phases.

\subsection{Data Processing}\label{sect:Sun_processing}
Initial analysis followed a similar approach to that of Paper 1, where the spectra for a single day were stacked after being continuum normalised and interpolated on to the wavelength grid of the low activity template spectrum. While the \logr~for the Sun is much lower than that of \acb, the large number of higher signal-to-noise (SNR) spectra technically allows greater sensitivity to spectral changes arising due to activity.

In our analysis of \acb, we employed the final merged and re-interpolated 1D-spectra data products from the HARPS data-reduction pipeline. However, in the case of the HARPS-N solar telescope, we observed what turned out to be etaloning of significant amplitude in the high SNR relative spectra (discussed in more detail in Section~\ref{sect:Ripple}). Initially, we thought the observed ripple pattern might be due to the interpolation to a constant wavelength grid done as part of the standard HARPS-N data reduction. We show later in Section~\ref{sect:Ripple} that this is not the case but is due to etaloning of an optical element within the light path of the instrument. Nonetheless, we used the 2D HARPS-N spectra (a lower-level data reduction pipeline product) that consist of stacks of individually extracted echelle orders prior to applying a wavelength solution and stitching into the 1D-spectra, adding some additional steps in the subsequent data processing not outlined in Paper 1.

First, the blaze profile was removed by dividing through by the blaze calibration file available per day, and if more than one blaze file was available these were averaged together. The blaze correction was applied on an order-by-order basis, where individual blaze functions are available for each order. Second, a wavelength solution was then applied to the 2D spectra using:
\begin{equation}
	\lambda = a_0 + \sum_{i=1}^{3 }a_i n_{pix}^i
\end{equation}

\noindent where $n_{pix}$ is the pixel number, and $a_{i}$ are calibration coefficients defined per echelle order and are determined as part of the standard HARPS-N data reduction pipeline. After this point the data were processed in a similar fashion to Paper~1; a master template spectrum was created to act as a reference wavelength grid, as well as being the low solar-activity comparison for the data. For this low-activity master template, data from $26^{\rm{th}}$ June 2016 was used, and the spectra from this day were individually checked for any obvious issues. A total of 107 spectra from this day were selected. The header files contain the SNR per order, which were used to identify the highest SNR spectrum to use as the template to continuum normalise all the other spectra. A mean of the SNR values for orders $40-50$ was used for the selection. Spectra from this day showed SNR values ranging from 267 to 357, with $>$90$\%$ of the spectra used for the creation of the low-activity master template having a SNR of $>$300.

Each echelle order was then RV corrected, but rather than using the barycentric RV defined by the HARPS pipeline, the RV corrections were calculated using the JPL Horizons system\footnote{JPL Horizons: \url{https://ssd.jpl.nasa.gov/horizons.cgi}}. While the HARPS-N pipeline is able to accurately measure RVs to the \ms{}~level, the instrument is not calibrated for taking solar data. The effects of the solar-system planets (most notably Jupiter) on the Sun are not properly corrected for when moving to this barycentric reference frame -- see \cite{CollierCameron2019} for more detail. Therefore, we used JPL Horizons to remove the planetary induced solar RVs as seen from the Earth following the procedure outlined in \cite{CollierCameron2019}. In this sense, the HARPS-N solar telescope dataset presents a unique testbed for studying astrophysical noise since it provides data on the only star for which we can confidently say that all the planetary RV signals have been removed. This means that, in our analysis of the impact of stellar activity, we do not have to disentangle its RV effects from those induced by bona-fide planets -- providing a `clean' dataset for such stellar activity studies.

The effects of differential extinction across the Solar disk were not considered as part of our data processing. This effect is due to the Solar surface being resolvable in the sky (i.e. it cannot be treated as a point source) causing different parts of its surface to experience different levels of extinction as the Sun moves across the sky. When summed together in a disk-integrated way (as is done with the HARPS-N solar telescope), this can affect the line profile shape, producing a signal similar to that seen for Rossiter-McLaughlin type measurements, where the leading, or trailing, hemisphere of the Sun would be differently extinguished. As can be seen later in Section~\ref{sect:Ripple}, the effects of differential extinction are not apparent in our results and that, although not directly considered, the effects are reduced as part of our data processing.

These blaze and RV corrected 2D spectra were next interpolated onto a common wavelength grid using a spline fit, with a spline knot defined at each point on the reference spectrum. Finally, all spectra were stacked per echelle order using a weighted mean with the weights being the same mean of the SNR in orders $40 - 50$ defined earlier. Unlike the 1D spectra, which interpolates each echelle order onto a set wavelength scale, we decided to preserve the original dispersion of each order. In this analysis we are searching for features within each echelle order, when comparing a single order over a range of observations this dispersion does not change and so will not affect our results. In addition, keeping the original dispersion helps limit any propagation of errors that might be caused by  re-interpolating multiple times.

For all high activity spectra, a similar process was followed but with a number of caveats. First, the low activity master templates created were used to define the wavelength grid and second, rather than a by-eye evaluation of the spectra, a cut of $>100$ was set in the SNR to remove bad spectra. Again, the spectra were RV corrected using the JPL Horizons data and stacked per echelle order using a weighted mean. Finally, each stacked spectral echelle order was divided by the equivalent low-activity master template echelle order to produce relative spectra for each order. On producing the relative spectra, however, a significant sinusoidal variation could be seen across the entire range, as mentioned earlier. The left panel of Figure~\ref{fig:Ripple} shows a few examples of the etaloning seen in the relative spectra, which was persistent across all the relative spectra, becoming more obvious towards redder wavelengths.

\subsection{Investigation and Removal of the Etaloning Pattern}\label{sect:Ripple}
An initial investigation of the periodic intensity fluctuations across each echelle order showed that these ripples do not have a constant period, but instead their period (measured in $\angstrom$) increases from the blue to the red end of the spectrum. The separation between peaks at $\sim$4000 $\angstrom$ is $\sim$0.2 $\angstrom$, which becomes $\sim$0.5 $\angstrom$ at $\sim$5800 $\angstrom$. The pattern also appears invariant to the activity of the Sun, with the variations of \logr~over the period range having no impact on the strength or periodicity of the ripple pattern. This dispersion and lack of correlation with changes on the Sun, coupled with observations of the same periodicity in flat-fielding images suggest that the source of the pattern is instrumental and is likely due to etaloning of an optical element. Below, however, we explore potential other sources (i.e. bad data and/or bad data processing) that could be responsible for the pattern. Our investigation of the etaloning focused on relative spectra generated for the most active phase studied in this paper, namely the $14^{\rm{th}}$ July 2016 (with the $26^{\rm{th}}$ June 2016 used as the low activity template comparison). These days were chosen as they show a large difference in \logr, and were expected to show the strongest pseudo-emission features in the relative spectrum if present. As the purpose of this investigation was to identify the source of, and remove, the etaloning (while maintaining any activity related pseudo-emission features), relative spectra showing larger pseudo-emission features were selected in order to aid the evaluation of the effectiveness of the removal process.

Prior to investigating the etaloning, each spectrum from the sample were individually checked for obvious features that could lead to the pattern. Effects from echelle order mismatch, poor seeing/weather, and RV offsets were considered as potential sources of the pattern but none of the spectra showed any effects from these. As production of the relative spectra involved continuum matching and interpolating the wavelength grid to that of the low-activity template, the choice of low activity template could also be the source of the pattern observed, or at least may potentially contribute to it. In order to check this, spectra from $12^{\rm{th}}$ November 2016 were selected as an alternative low-activity template but, again, the etaloning persisted.

\begin{figure}
	\includegraphics[width=\columnwidth]{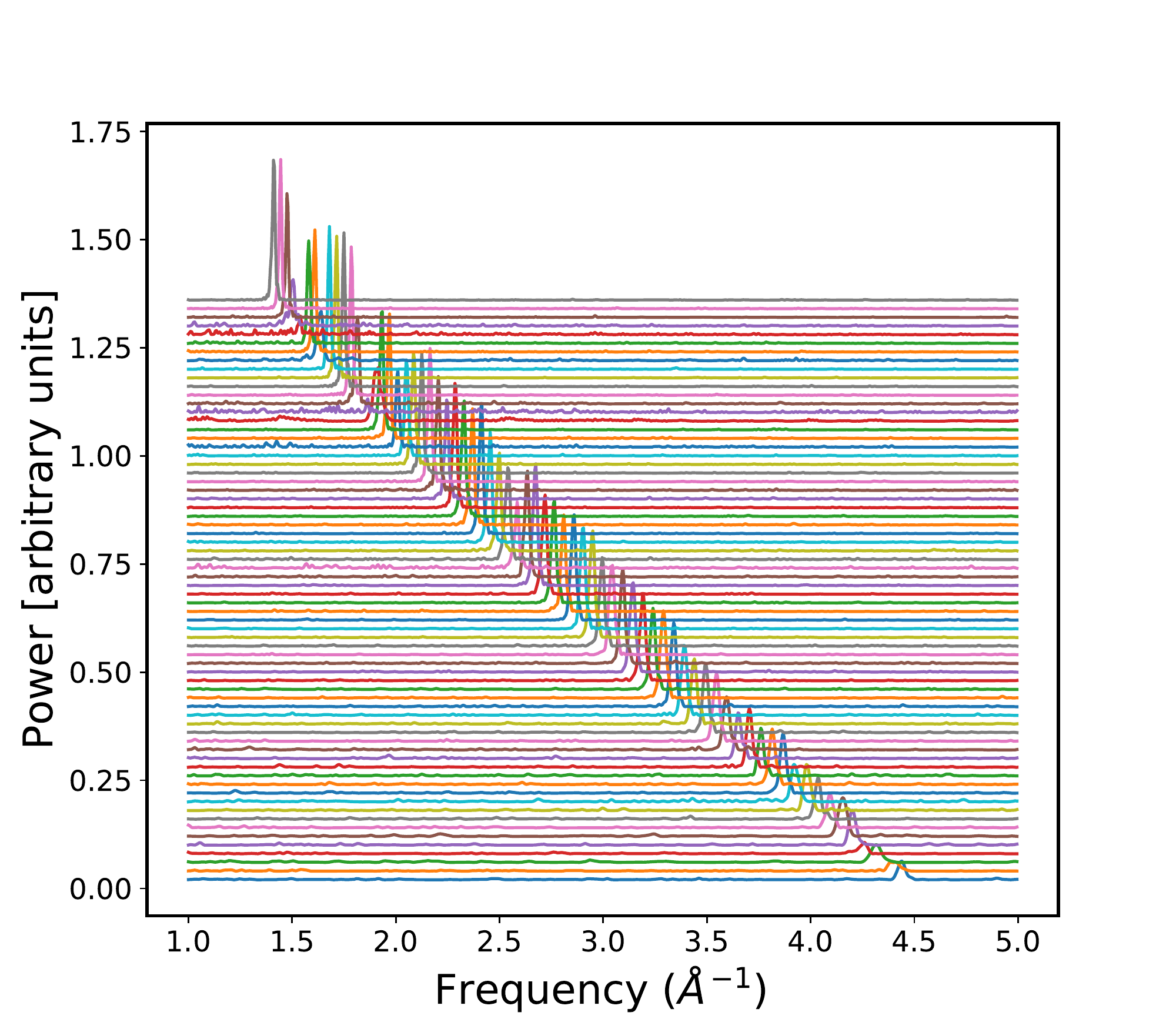}
		\caption[GLS periodogram of solar spectrum Echelle Orders]{The GLS per echelle order, with internal order 1 (central wavelength $\sim$3897\AA) at the bottom and internal order 68 (central wavelength $\sim$6797\AA) at the top highlighting the period of the ripple pattern. Per order, the ripple is close to having a non-evolving static periodicity.}
	\label{fig:GLS_Sun}
\end{figure}

As outlined above, instead of working with full 1D spectra, we used 2D spectra to remove any errors introduced by resampling of the data to an even wavelength grid. The data were processed following the procedure outlined earlier (Section \ref{sect:Sun_processing}) to produce relative spectra for each order. The etaloning pattern was analysed for each order using a Generalised Lomb-Scargle(GLS) periodogram \citep{Zechmeister2009}, which showed that the frequency (measured in $\angstrom^{-1}$)  with peak amplitude decreases moving from the lower echelle orders to the higher orders (see Figure~\ref{fig:GLS_Sun}). A number of individual orders from other daily stacked relative spectra were also analysed using a GLS periodogram. These showed a similar result, with equivalent echelle orders showing a peak in power at the same frequency. Importantly, the GLS periodograms show that, per order, the ripple pattern is well defined by a small range of frequencies.

\begin{figure}
	\includegraphics[width=\columnwidth]{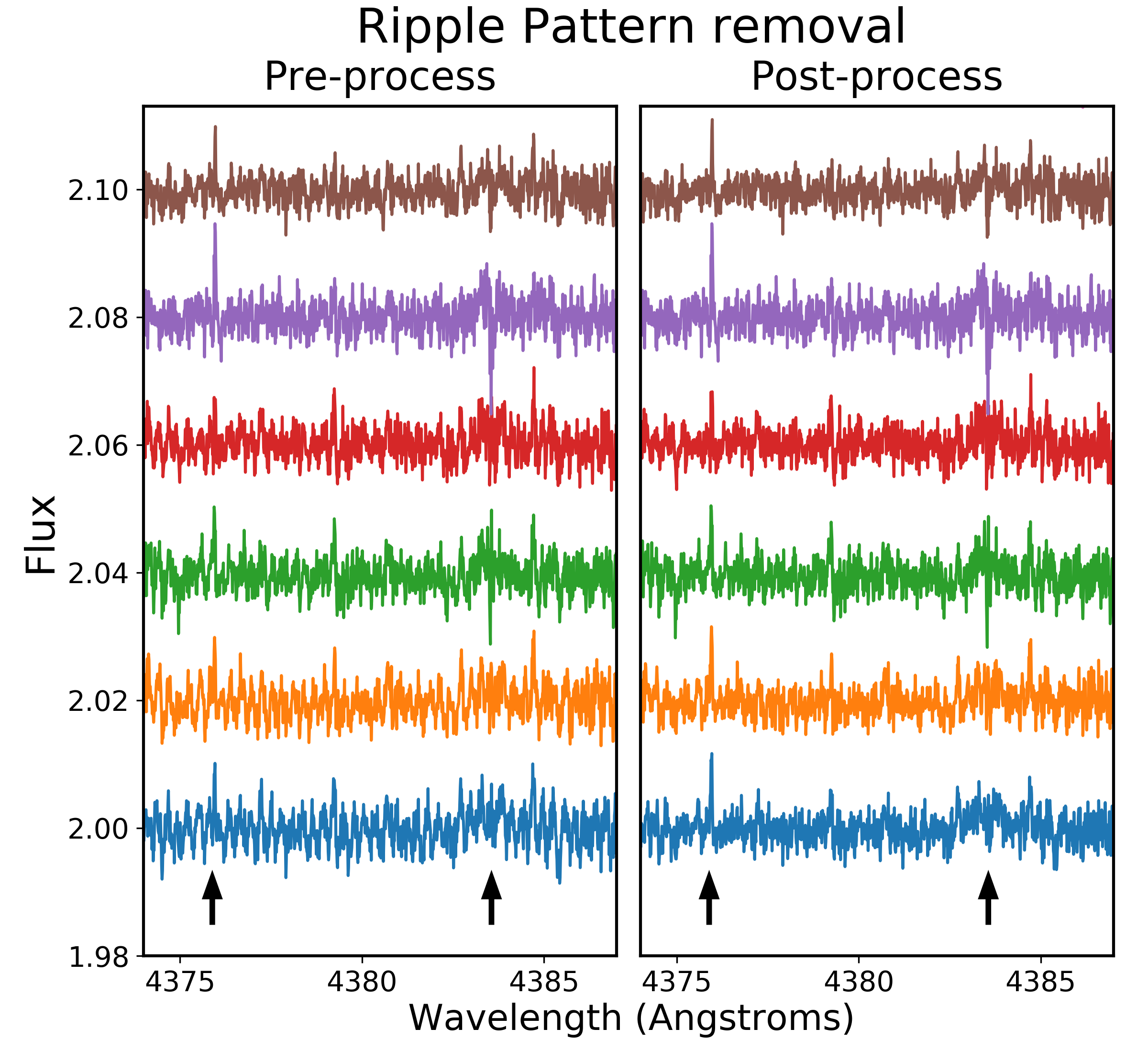}
		\caption[Ripple Removal Process]{Pre- (left panel) and post-processing (right panel) of the daily stacked relative spectra to remove the etaloning. Each relative spectrum, produced by dividing each spectrum by the low-activity template, is from a different day and shows the variation in the strength of the pattern. This demonstrates the need for a manual by-eye reduction. The black arrows indicate the position of known features in the relative spectrum (see Section \ref{sect:features} where we show that these are indeed activity-related features). The relative spectrum in blue shows a good example of the success of the etaloning removal process, where the features cannot be clearly seen prior to the removal but are an obvious signal in the post-processed relative spectrum.}
	\label{fig:Ripple}
\end{figure}

This evidence points towards the source of the pattern being the etaloning of an optical element. Per echelle order the pattern is well constrained to a small frequency range -- making it a relatively clean signal for the purposes of removing it. The slowly evolving frequency of the etaloning in each individual order can be approximated by fitting a number of sinusoids to the relative spectra. Over a single echelle order the etaloning was captured by applying a low number of sinusoidal fits, with the fit done on an order-by-order basis. To start, a GLS periodogram was performed on the relative spectra of a single echelle order, the highest peak was then identified and set as the frequency of the sine wave -- with the amplitude of the fit calculated following the derivation from \cite{Zechmeister2009}. Finally, a  Markov Chain Monte Carlo (MCMC) simulation was used to obtain the wave phase.

The MCMC approach is typically used when analysing models with multi-dimensional distributions helping prevent these models fits getting stuck in local maxima/minima. For this analysis, the MCMC was chosen to model the phase of the sinusoidal fit as it was found that more conventional $\chi^{2}$ minimisation techniques were not effective at determining a precise value for the phase. This is due to the weak nature of the signal, as well as the model only approximating the evolving frequency of the pattern. Rather than a full rigorous MCMC approach, we use the method to help better explore the parameter space around the best fit value. This MCMC-like approach uses similar acceptance/rejection criteria but we did not go through the process of checking for convergence or mixing. As we are fitting only one parameter strong correlation between parameters was not an issue and we found that the technique converged very quickly (within a few 100 steps from a any random starting position). The MCMC-like approach was run with multiple, sequential burn-in phases, after which a final MCMC was run to explore the local parameter space. A boundary condition was also put on the phase that only allowed the MCMC to uniformly sample values between 0 and $2\pi$. A total of 10 burn-in phases, each having 1000 steps, were run one after the other. At each step the $\chi^{2}$ value of the fit was calculated and compared to the previous fit for acceptance or rejection based on the Metropolis-Hastings algorithm \citep{Metropolis1953}. At the end of each burn-in run the step size was adjusted and taken as the standard deviation of the values in that run. This allows for a larger step size when far from the best fit and a reduced step size as the model parameters approach good estimates. The starting position of the next burn-in was taken as the last value of the previous run (with the phase initially set to zero prior to the first burn-in run). This burn-in phase is used to get good estimates of the parameter starting position and step sizes. A final MCMC was run over 5000 steps and the values of this run were used to estimate the phase. A histogram of this final MCMC run was calculated and a Gaussian was fit to it, where the peak of the fitted Gaussian defined the phase of the sine wave model.

This model sine wave was then removed from the relative spectrum and the residuals checked, using a GLS periodogram, to determine its effectiveness. If any strong periods are found in the periodogram, the process was rerun over the residuals to fit the next strongest periodic signal. This process is repeated until the GLS shows no strong peaks above the noise floor. It was found that by applying the etaloning removal process 3 times the evolving periodicity was almost entirely removed. On some occasions only 1 or 2 sine waves were needed to remove the etaloning. Conversely, by fitting additional sine waves, variations were re-introduced into the relative spectra. The choice of the final model then required a by-eye approach to stop any overfitting. Examples of the etaloning and the results after its removal are shown in Figure~\ref{fig:Ripple}, where the pattern can be seen in some cases to completely obscure real spectral features.

As a check, each of the echelle orders from the relative spectra of $14^{\rm{th}}$ July 2016 were processed to remove the etaloning (i.e. the same date initially chosen for the investigation into the etaloning). While the approach used was able to substantially reduce the impact of the etaloning per echelle orders, it was found that for a number of orders the residual effects of the etaloning still dominated the relative spectra. However, we were able to identify 8 out of the 69 orders (see Table \ref{table:Orders}) that exhibited strong features that could be activity-related spectral variations similar to those of \acb~reported in Paper~1. In addition, the $14^{\rm{th}}$ July 2016 represents the highest activity level (as measured by \logr) of the 2016 interval, meaning that any activity-based features would be strongest here. For these reason, only the 8 orders in Table~\ref{table:Orders} were processed to remove the etaloning for the rest of the 2016 interval. This has the added advantage of speeding up the analysis as the etaloning removal process is time intensive, requiring a visual inspection of each model fit to confirm its effectiveness.

\begin{table}
    \centering
    \begin{tabular}{c|c}
        \hline
		\hline
        Internal    & Wavelength \\
        Order       & Range \\
		\hline 
        15  & $\sim 4253 - 4300$ \\
        18  & $\sim 4344 - 4393$ \\
        22  & $\sim 4472 - 4522$ \\
        24  & $\sim 4538 - 4589$ \\
        30  & $\sim 4751 - 4804$ \\
        36  & $\sim 4985 - 5041$ \\
        37  & $\sim 5026 - 5082$ \\
        61  & $\sim 6269 - 6339$ \\
        \hline
    \end{tabular}
    \caption{List of the 8 internal echelle orders selected for processing to remove the etaloning. None of the other 61 orders showed any pseudo-emission features and so where not considered for the rest of the analysis. Note that the wavelength range is only an approximate value due to RV variations changing the wavelength window in each orders.}
    \label{table:Orders}
\end{table}

\section{Results}\label{sect:features}
\begin{figure*}
	\includegraphics[width=\textwidth]{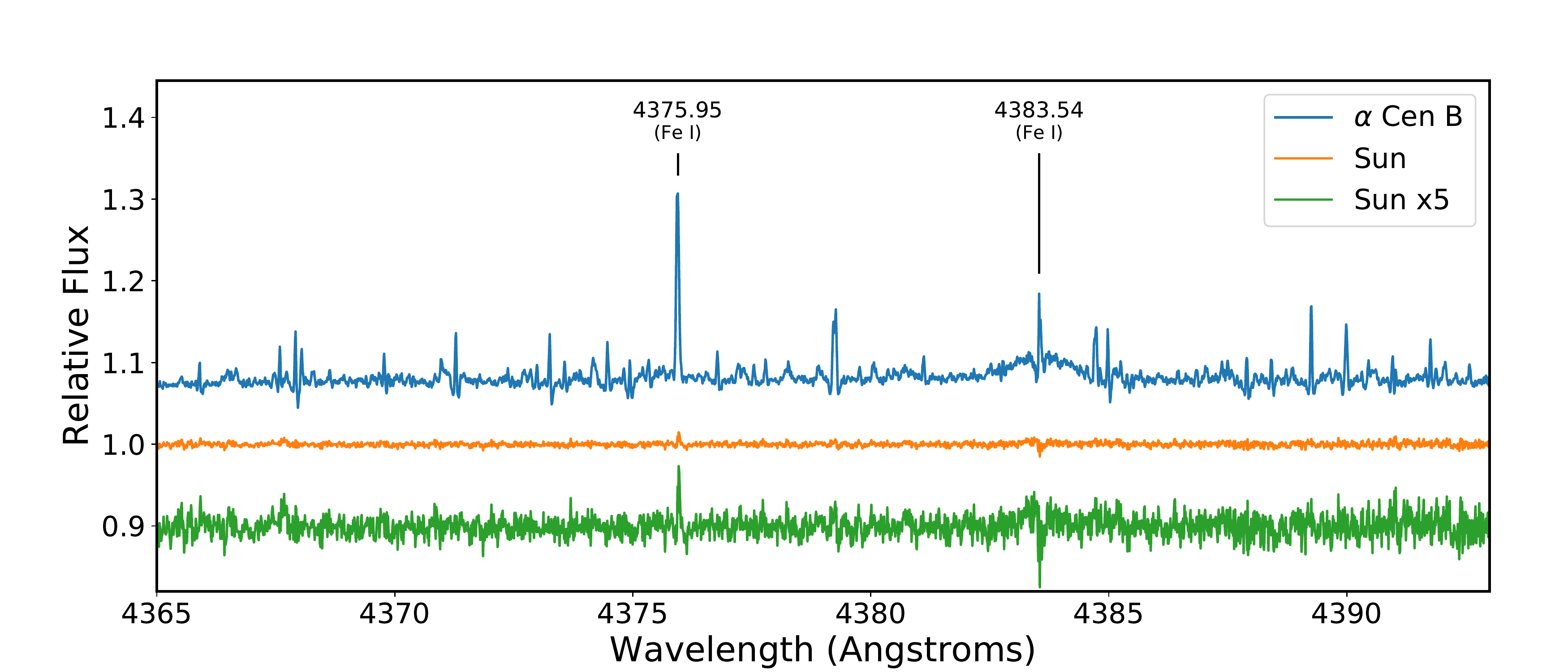}
		\caption[Relative spectrum of the Sun versus \acb]{Comparison of the features found on the relative spectra (generated by dividing a high-activity spectrum by a low-activity template) of \acb~(blue), discussed in Paper~1, with features found on the Sun (orange). The strongest narrow peak belongs to the \ion{Fe}{I} $4375 \angstrom$ line, and the broader feature is due to \ion{Fe}{I} $4383 \angstrom$ as indicated. The lower activity level of the Sun means that the features are considerably weaker. Both the orange and green lines show relative spectra from the Sun, but with the values increased by a factor of 5 for the green line, allowing the features to be more easily compared. Note that, for the Sun, only a few features are easily visible compared to the forest of features found on \acb, which is due, in part, to residual etaloning noise.}
	\label{fig:Sun_acb_compare}
\end{figure*}

The relative spectra of the final 8 reduced echelle orders all showed some indications of features that could be due to solar activity. The three strongest features (out of the hundreds present) highlighted in our study of \acb~in Paper~1 ($4375 \angstrom$, $4383 \angstrom$, and $4404 \angstrom$) are also seen in the Sun, but are far weaker due to the Sun's lower activity level. Figure \ref{fig:Sun_acb_compare} shows a comparison between the relative spectrum of the Sun (this study) and that of \acb~(from Paper~1) for echelle order 18 -- which contains prominent pseudo-emission features arising in the  $4375 \angstrom$ and $4383 \angstrom$ lines (the $4404 \angstrom$ line, although present, is not shown as this line is in a different echelle order). Both of these features are also clearly visible in the Sun, while the multitude of weaker features seen in the relative spectra of \acb~are less easy to identify visually in the solar data.

As a note, in Section \ref{sect:data} we mentioned that the effects of differential extinction were not considered, and that this could impact the shape of the observed solar-absorption lines -- mimicking a Rossiter-McLaughlin like signature as the leading or trailing hemisphere becomes more heavily extinguished. However, to first order, such an effect would impact all lines equally, which is not what we see here. In addition, our use of daily-stacks would also act to largely cancel out such intra-day effects. Thus, the impact of differential extinction is not responsible for the features seen in Figure \ref{fig:Sun_acb_compare} and, as we explain later, such features are clearly linked to the solar rotation period and appearance of active regions. This also provides confidence that such features do not arise due to incorrect removal of the etaloning (Section~\ref{sect:Ripple}).

Visual inspection of the relative spectra was used to identify features that showed variability that could be linked to solar activity -- from this, 6 relatively strong features were identified for further investigation (see Table~\ref{table:Sun_lines} and Appendix A for plots of each of the features found). Other features initially identified were either too weak to be seen in many of the other relative spectra, or did not show variability consistent with activity (and are likely weak telluric lines that are only seen due to the very high signal-to-noise of the daily stacked spectra). A Gaussian was fit to each of the activity-related features identified, with the area of the fitted Gaussian taken as an estimate of the feature strength (a pseudo-equivalent width measure similar to that used in Paper~1). For cases where the features were too weak to measure, no pseudo-equivalent width was calculated.

\begin{table*}
	\centering
		\begin{tabular}{cccccccccccc}
			\hline
			\hline
			Line			&	Element/	 	&	Echelle	&	&&	Pearson's r	&	\\
			(wavelength $\angstrom$)	&	Ion		&	Order	&	Spots	&	faculae	&	$\log{\ R^{\prime}_{HK}}$ & H$\alpha$ & H$\beta$ & H$\gamma$ & H$\delta$ & He & Na\\
			\\[-1em]
			\hline 
			4375.95		& \ion{Fe}{I}	    & 18	& 0.214 &	0.392 & 0.387 & 0.105 & -0.133 & 0.272& 0.099 & 0.142 & 0.315\\
			4501.35		& \ion{Ti}{II}      & 22	& 0.651 &	0.731 & 0.740 & 0.069 & -0.211 & -0.185 & 0.150 & -0.192 & -0.006\\
			4783.43 	& \ion{Mn}{I}       & 30	& 0.209 &	0.496 & 0.454 & 0.114 & -0.335 & -0.170 & 0.012 & 0.134 & 0.334\\
			5007.27		& \ion{Fe}{I}	    & 36	& 0.118 &	0.387 & 0.546 & 0.101 & 0.091 & -0.130 & 0.213 & -0.123 & -0.081\\
			5012.12		& \ion{Fe}{I}	    & 36	& 0.165 &	0.495 & 0.531 & 0.254 & -0.307 & -0.123 & 0.277 & 0.288 & 0.438\\
			5014.27		& \ion{Ti}{I}	    & 36	& 0.340 &	0.684 & 0.587 & 0.029 & -0.417 & -0.062 & 0.187 & 0.016 & 0.288\\
			\hline
		\end{tabular}
		\caption[Pearson's r correlation of relative features (2016 interval)]{Lines that show a feature in the relative spectra and the echelle order each line is in. The Pearson's r statistic is shown for each of the features with respect to spot filling factor, faculae filling factor, \logr, and the daily medians of the activity indexes measured by \cite{Maldonado2019} (see text for details). This is for the interval $16^{\rm{th}}$ June to $24^{\rm{th}}$ August 2016. The identified element/ion is taken from the Vienna Atomic line Database \citep[VALD,][]{Ryabchikova2015}. }
		\label{table:Sun_lines}	
\end{table*}

\begin{figure}
	\includegraphics[width=\columnwidth]{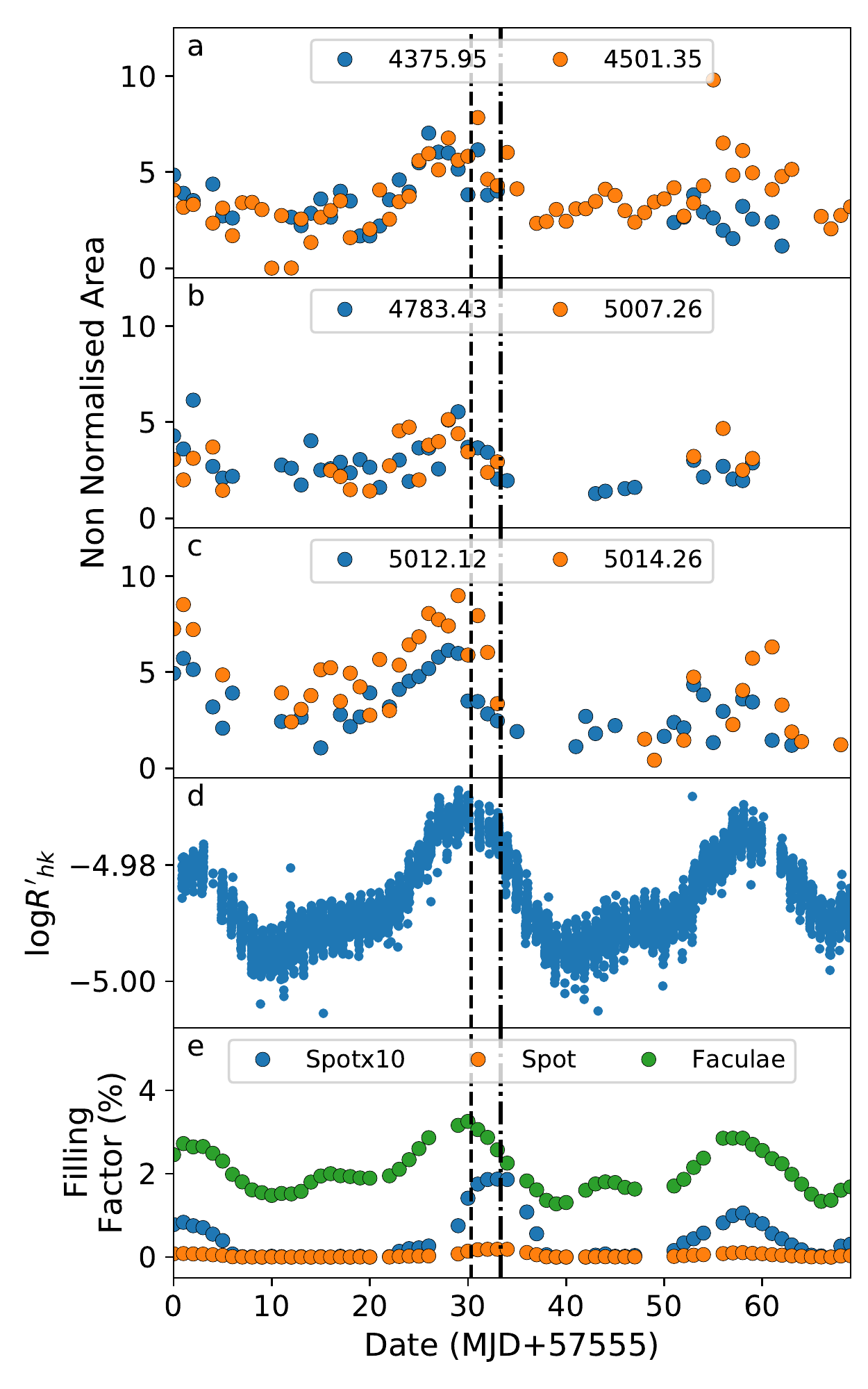}
		\caption[Equivalent Width of relative features on the Sun (2016 interval)]{Pseudo-equivalent widths for all 6 features separated over 3 plots for ease of viewing. (a) features at $4375 \angstrom$ and $4501 \angstrom$, (b) features at $4783 \angstrom$ and $5007 \angstrom$, and (c) features at $5012 \angstrom$ and $5014 \angstrom$ (missing data points are due to the feature at that time being too weak to reliably measure). These are compared to \logr~(d) and spot and facular filling factors (e). The spot filling factor has been increased by a factor of 10 to allow ease of viewing. The dashed and dot-dashed lines show the peak in faculae and spot filling factors, respectively. Note the faculae have a phase offset in their peak filling factor, indicating unassociated facular regions, and how the changing equivalent width of the pseudo-emission features appear to better trace the variation in faculae filling factor rather than spots filling factor.}
	\label{fig:SunComp}
\end{figure}

The pseudo-equivalent widths of the features were compared to \logr~(as was done in Paper~1) to determine if activity is the driving force behind the changes in strength observed.  While \logr~is a good proxy for the level of activity on the surface of a star, it traces chromospheric activity and not changes to the photosphere. In the Solar case, we have the additional benefit of images of the surface. Changes in the relative spectra can be directly compared to inhomogeneities seen on the photosphere of the Sun and additionally allow for spot and faculae contributions to be monitored separately. We use the method outlined in \cite{Haywood2016}, where the authors used images from the Helioseismic and Magnetic Imager (HMI) aboard the Solar Dynamics Observatory \citep[SDO,][]{Scherrer2012} to measure the filling factors of facular regions and spots on the visible surface of the Sun. Using SDO/HMI flattened intensitygrams, line-of-sight Dopplergrams and unsigned magnetograms, a threshold in the magnetic field strength of each pixel ($B > 24$G) was used to separate quiet Sun from active regions, while a threshold in the flattened continuum intensity ($I > 0.89 \hat{I}_{QS}$, where $\hat{I}_{QS}$ is the mean intensity of the quiet Sun) allows for spots and faculae to be differentiated. The pseudo-equivalent width of the features in the relative spectra were then compared to \logr~and the spot and faculae filling factors as shown in Figure~\ref{fig:SunComp}. 

All 6 features show variations that appear to correlate with \logr, showing a peak in the pseudo-equivalent width measurements close to where the peak in \logr~can be seen. This is most clear for the lines at $4375 \angstrom$ and $4501 \angstrom$ in Figure~\ref{fig:SunComp}, where the two peaks in \logr~at $\sim 30$ days and $\sim 57$ days have equivalent peaks in the pseudo-equivalent width of the two lines. This is a similar result to that seen in Paper~1 and adds further evidence that the features are generated by activity. Additionally, comparing the pseudo-equivalent widths with the spot and faculae filling factors shows an offset in the peak of the pseudo-equivalent widths compared to the peak in the spot filling factor. Pearson's r correlations for pseudo-equivalent widths versus spot and faculae filling factors, as well as \logr,~are shown in Table~\ref{table:Sun_lines}. This indicates that the pseudo-emission features being tracked in the solar spectra change in a manner more strongly correlated to changes in the faculae filling factor, rather than the spot filling factor.

\cite{Maldonado2019} conducted a study of conventional solar-activity indicators (using the same HARPS-N solar telescope data used in this work) and include activity indexes of the H$\alpha$, H$\beta$, H$\gamma$, H$\delta$, Na {\sc{i}} D1 D2, and He {\sc{i}} D3 lines. We also present the Pearson's r correlation coefficients between the daily medians of these indexes and our pseudo-equivalent widths in the final 6 columns of Table~\ref{table:Sun_lines}. We find no particularly strong correlation or anti-correlation between our pseudo-emission features and these other activity indicators, with the strongest correlations still with the faculae filling-factors and \logr. The lack of correlation with these other activity-indicators is perhaps unsurprising. \cite{Maldonado2019} note that measurements of He {\sc{i}}, Na {\sc{i}}, and H$\alpha$ are all impacted by telluric lines either contaminating the cores or continua of these indexes. We also checked the spectra over the dates analysed, and indeed found substantial telluric contamination of these lines during the dates under study. The pseudo-emission features, however, are telluric free - demonstrating a potential advantage of using these new activity indicators. In addition, the response of the H$\alpha$ line to activity can be complex, first showing a deeper absorption profile as activity increases until the electron density becomes sufficiently high that the line then becomes collisionally excited, at which point increasing core emission is observed. Furthermore, on short timescales it is known that \logr and H$\alpha$ does not correlate strongly. Indeed, \cite{Maldonado2019} find that the Pearson's r correlation coefficient between H$\alpha$ and S-index for the Sun can vary from nearly -0.7 to +0.6 when analysing different $\sim$90-day observation windows, and for other stars a large dispersion is also seen in the correlations between H$\alpha$ and the Ca {\sc{ii}} H \& K activity indicators (e.g. \citealt{Cincunegui2007}). This dispersion may be caused by filaments, which are more prominent in the Balmer lines (e.g \citealt{Meunier2009}). Thus, if filaments are not well correlated with faculae, then such activity-indicators may fail to follow the faculae filling factors.

\begin{figure}
		\begin{center}
		\includegraphics[width=\columnwidth]{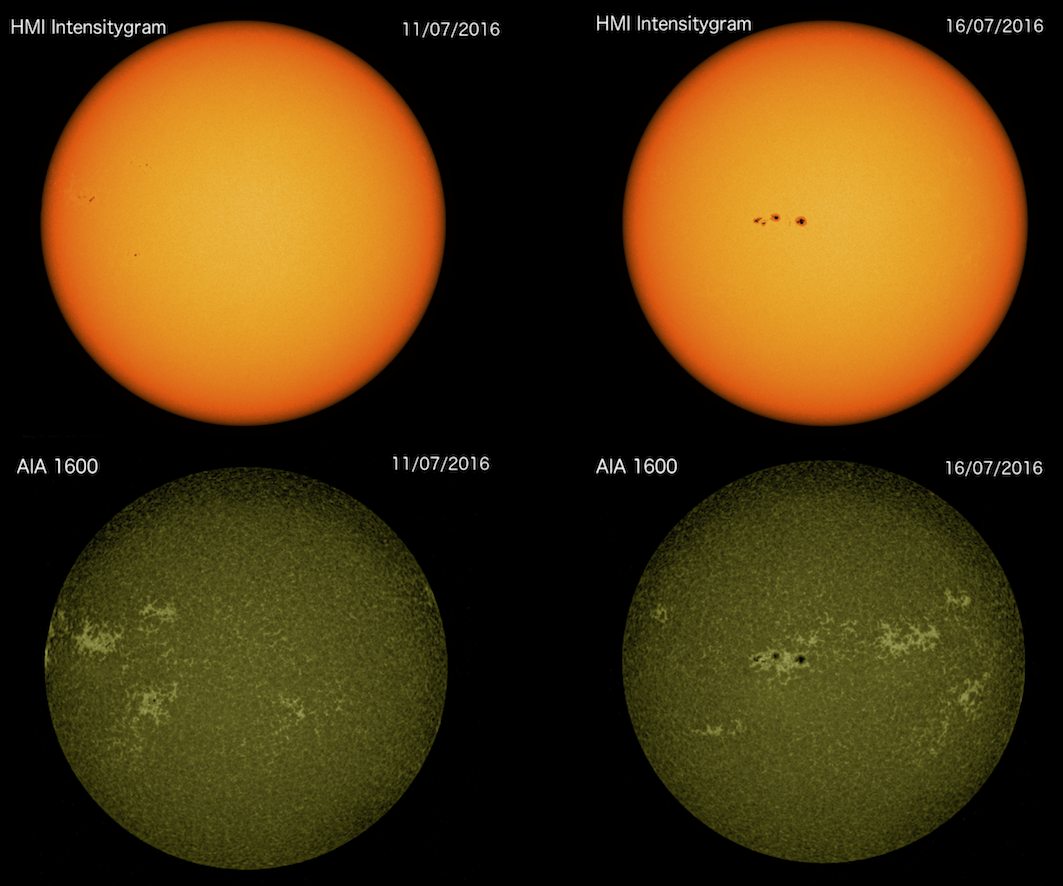}
			\caption[HMI Solar images (2016 period)]{SDO/HMI images of the Sun from two different dates highlighting the extended unassociated facular regions. The images show two filters chosen to better view each type of activity, with the spots clearly seen in the HMI Intensitygram (top panels) and the faculae most visible in the AIA 1600 (bottom panels). The left panels (taken from the $11^{\rm{th}}$ July 2016) clearly shows that faculae, with little to no associated spots, rotates into view first. A few days later on $16^{\rm{th}}$ July 2016 (right panels) an extended spot group then rotates into view and is offset from the major facular region.}
		\label{fig:SunHMI}
		\end{center}
	\end{figure}

To further test the correlation between the pseudo-emission features in the solar spectra and the faculae filling factor, the interval of the $4^{\rm{th}}$ to $27^{\rm{th}}$ July 2016 was isolated in the dataset. This period is of interest as a facular region rotates onto the visible hemisphere several days prior to the emergence of a prominent spot group. This provides a small time window where we can directly disentangle whether the spectral variability we observe is driven by faculae or by spots. This lag between the faculae and the spots rotating into view is clearly demonstrated in Figure~\ref{fig:SunHMI}. This lag can also be seen in the bottom panel of Figure~\ref{fig:SunComp}, where the faculae filling factor can be seen to rise several days before the spot filling factor follows suit - with the peaks in the respective filling factors indicated by the vertical dashed/dot-dashed lines. As can be easily seen, the pseudo-emission feature strengths closely follow the faculae filling factor. Pearson's r correlation coefficients were measured for the features over this time range, and shows an overall increase in the correlation with the faculae, but a slight decrease in correlation with the spots (see Table~\ref{table:Sub_range}) -- further supporting our interpretation that the pseudo-emission features are faculae driven. This was also what we concluded in Paper~1, where we used simple modelling to fit the morphologies of the features seen in \acb, which consisted of a limb-darkened disk (representing the stellar surface) with an active region (either spot or faculae) placed at disk centre. The properties of the disk and active region could then be altered to fit the different pseudo-emission morphologies observed (more details can be found in Paper~1). The results of this model showed evidence that faculae is needed to fully account for the pseudo-absorption troughs observed. This result further corroborates our findings in Paper~1. 

	\begin{table*}
		\centering
			\begin{tabular}{cccccccccccc}
				\hline
				\hline
				Line			&	Element/	 	&	Echelle	&	&	Pearson's r & \\
				(wavelength $\angstrom$)	&	Ion		&	Order	&	Spots	&	faculae & $\log{\ R^{\prime}_{HK}}$ & H$\alpha$ & H$\beta$ & H$\gamma$ & H$\delta$ & He & Na \\
				\\[-1em]
				\hline
				4375.95		& \ion{Fe}{I}		& 18	& 0.252     & 0.646 & 0.768 & 0.233 & -0.068 & -0.198 & 0.365 & -0.119 & 0.151 \\
				4501.35		& \ion{Ti}{II}      & 22	& 0.635     & 0.832 & 0.838 & -0.031 & -0.099 & -0.003 & 0.193 & -0.193 & 0.014 \\
				4783.43 	& \ion{Mn}{I}	   	& 30	& -0.010    & 0.662  & 0.607 & -0.354 & -0.172 & -0.025 & -0.032 & -0.301 & -0.065 \\
				5007.27		& \ion{Fe}{I}		& 36	& 0.049     & 0.385 & 0.548 & 0.026 & -0.041 & -0.190 & 0.267 & -0.088 & 0.071 \\
				5012.12		& \ion{Fe}{I}		& 36	& -0.239    & 0.402  & 0.675 & 0.203 & -0.325 & -0.370 & 0.261 & 0.258 & 0.381 \\
				5014.27		& \ion{Ti}{I}		& 36	& 0.217     & 0.743 & 0.731 & -0.103 & -0.203 & -0.214 & 0.187 & -0.200 & 0.005 \\		
				\hline
	
			\end{tabular}
			\caption{Same as Table \ref{table:Sun_lines}, but for the range $4^{\rm{th}}$ to $27^{\rm{th}}$ July 2016, which is dominated by faculae that is unassociated with spots. The results indicate that the features better trace the changes in faculae coverage than spot coverage.}
			\label{table:Sub_range}	
	\end{table*}

\begin{figure}
		\includegraphics[width=\columnwidth]{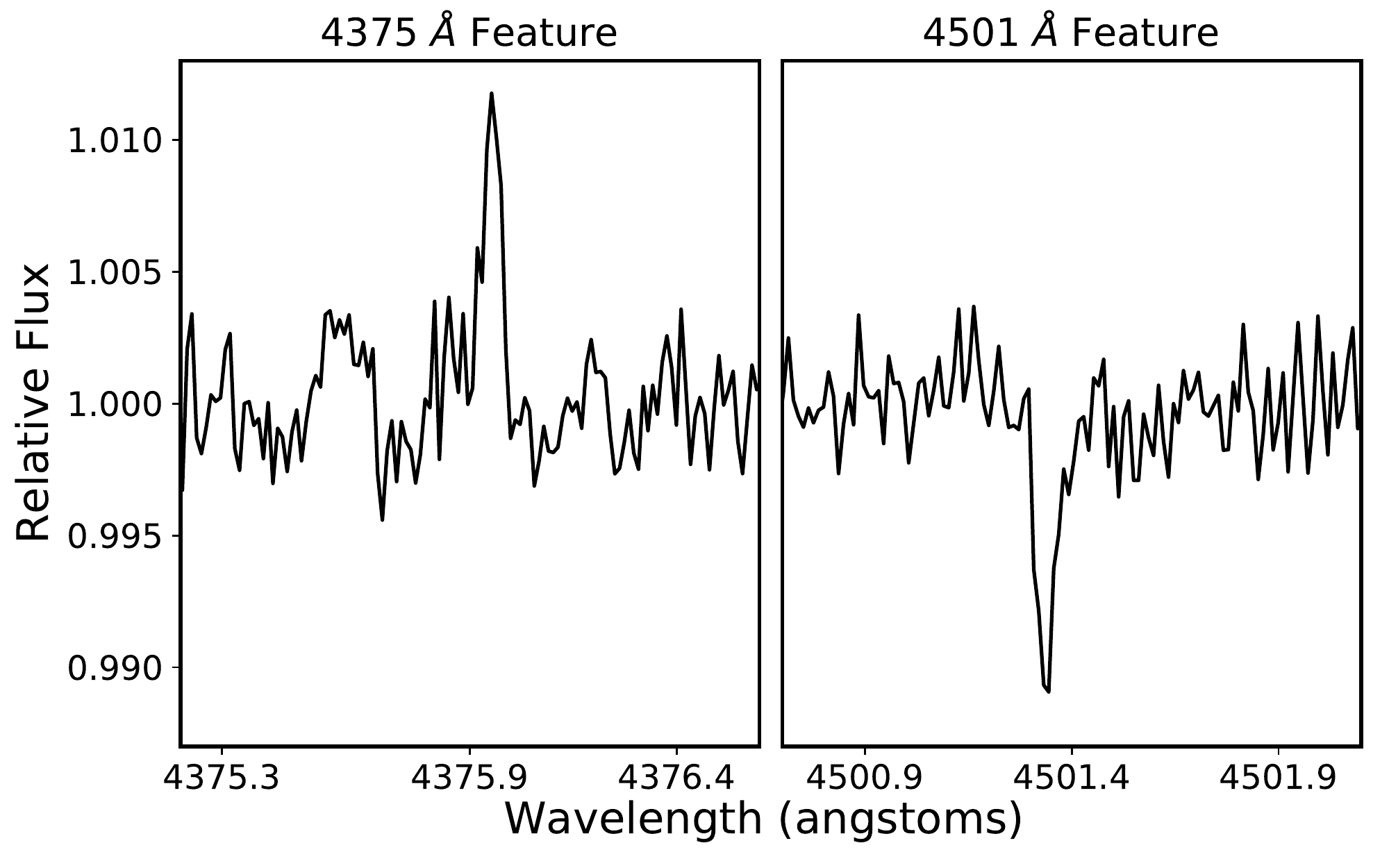}
			\caption[Example of $4501\angstrom$ pseudo-absorption dip]{Example of the pseudo-absorption dip of the $4501 \angstrom$ feature as well as the $4375 \angstrom$ feature for comparison. This example is from the $14^{\rm{th}}$ July 2016 and is one of the highest activity days of the 2016 interval.}
		\label{fig:4501_dip}
	\end{figure}

The feature at $4501.35 \angstrom$ does not show much of a change in correlation strength between the full 2016 interval dataset and the reduced set. This feature, however, is the only one to show a pseudo-absorption dip rather than a pseudo-emission peak in the relative spectra (Figure~\ref{fig:4501_dip} shows an example of this feature, as well as the 4375 $\angstrom$ feature for comparison). This suggests that the line responsible for this feature is deeper when the Sun is in a higher activity state, as opposed to the rest of the lines, which show pseudo-emission peaks and are shallower in the high activity state, and may indicate a different mechanism that alters this line. A more in-depth analysis of the different morphologies of the pseudo-emission lines (and the physical mechanisms responsible for these) will be the subject of a future paper.

Recent work by \cite{Milbourne2019} performed a complementary analysis to that presented here. They compare HARPS-N solar telescope RV and \logr~measurements to activity filling factors using a similar approach  as \cite{Haywood2016} to identify activity from SDO/HMI images. In addition to defining spot and faculae filling factors, as we do in this work, the authors further defined a network filling factor. Network is similar to faculae and differs only in size, \cite{Meunier2010} define network as being bright structures outside of active regions (while faculae are bright structures in active regions). Following this approach, \cite{Milbourne2019} split their bright activity filling factor into faculae (bright active regions having a size $>$20 micro-hemispheres) and network ($<$20 micro-hemispheres) filling factors.

We measured the correlation of the pseudo-emission features with respect to both the faculae and network filling factors as defined in \cite{Milbourne2019} and find that the features more strongly correlate to faculae than to network. This is not an unexpected result and can be understood by examining the nature of the two activity structures. Although both faculae and network cause brightening of the photosphere due to increased concentration of magnetic field lines, network is much more homogeneously distributed than faculae. This means that, over the short period range considered in this analysis, only a small change in the network filling factor is seen as the network does not evolve much over this timescale. Faculae, being more localised on the solar surface, shows a much larger variation as extended active regions rotate on and off the visible hemisphere of the Sun. When correlating the pseudo-emission features presented in this work with faculae and network filling factors of \cite{Milbourne2019}, we see higher correlation with the more localised, rotationally modulated, faculae compared to the much more homogeneously distributed network. This comparison further strengthens the result that the features are not only driven by magnetic activity, but are tracking changes in the faculae filling factor.

\subsection{Radial velocity variations of features}
\begin{figure}
		\includegraphics[width=\columnwidth]{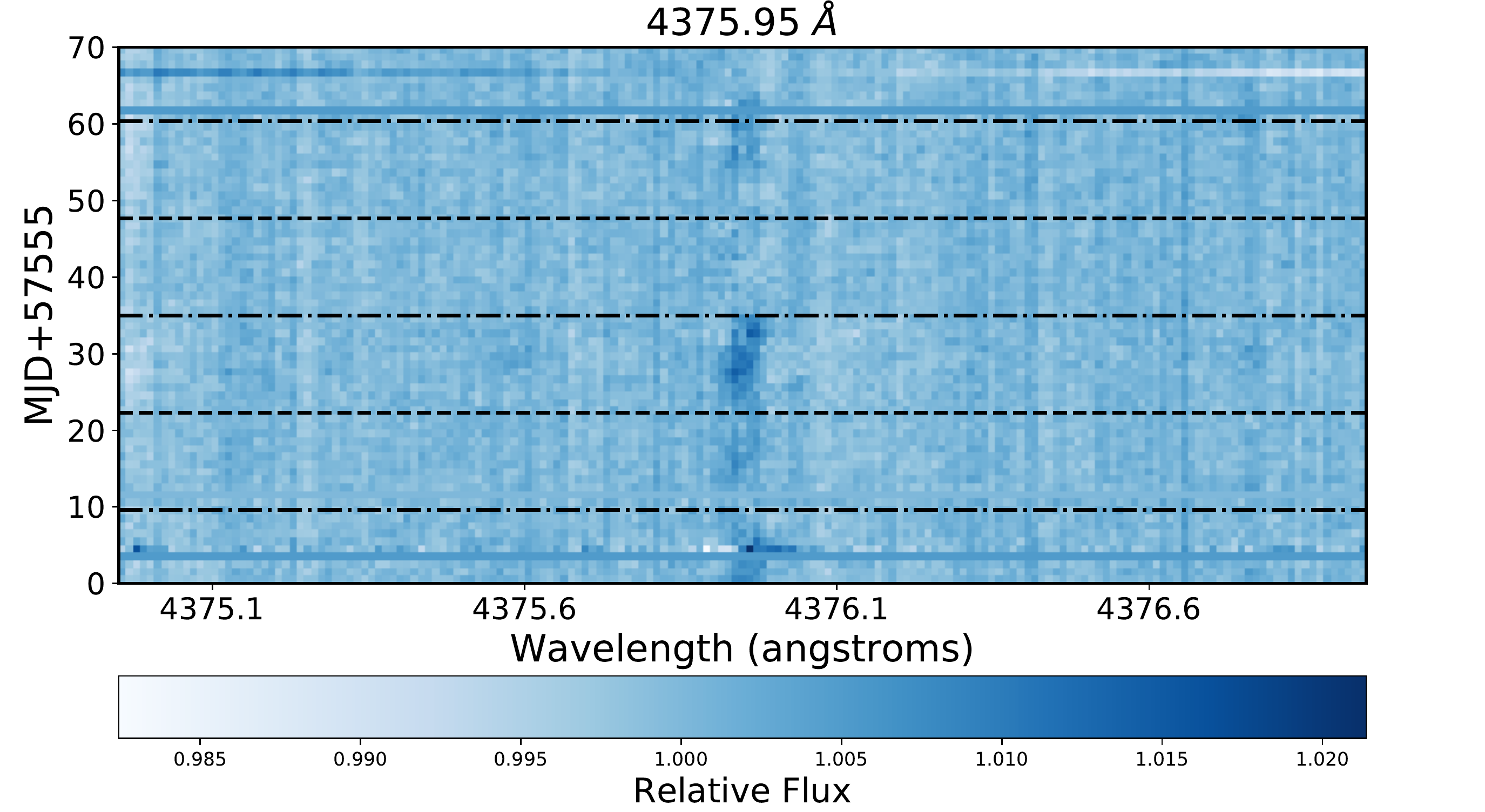}
			\caption[Trailed relative spectrum of $4375 \angstrom$ relative feature ]{Trailed relative spectrum of the $4375 \angstrom$ line, where the horizontal dashed and dot-dashed lines are the Carrington rotation period and half period, respectively. The main feature from the $4375 \angstrom$ pseudo-emission peak can be seen to have a positive slant indicating that the feature has a radial velocity component appearing to move from blueshift to redshift as it traverse the solar disk. Note the spectrum at $\sim{5}$ days has low signal-to-noise, explaining the offset feature. Featureless horizontal blue bands (e.g. as seen at day 4) represent periods where no HARPS-N solar telescope data was taken.}
		\label{fig:spec_map4375}
	\end{figure}

In order to investigate any RV motion the features may have, we create trailed spectra centred at the wavelength of each of the features. Figure \ref{fig:spec_map4375} shows a trailed spectrum of the strong pseudo-emission feature seen in the relative spectra for the $4375 \angstrom$ line. The dashed and dot-dashed lines represent the Carrington rotation period \citep[i.e. 25.38 days, ][]{Carrington1859} and half of that, respectively. As only one hemisphere of the Sun can be seen, the half rotation period can be useful in determining when an active region would migrate off the visible surface of the Sun after it first rotates into view. The position of the markers were chosen from the SDO/HMI images, with the zero-point based on identifying a large feature seen to rotate onto the visible surface of the Sun and is \textit{not} chosen to intentionally line up with any features in the relative spectra. In Figure \ref{fig:spec_map4375}, the feature at $4375 \angstrom$ is seen to be encapsulated within a half-rotation of the Sun, as one would expect from an active region rotating on and off the visible solar hemisphere. This can most easily be seen at $\sim$22 - 35~days, where the feature can be easily picked out from the background noise. The feature can also be seen to have a directionality showing a positive slant indicating a radial velocity component of the feature moving from blue-shift to red-shift, which would be expected if the feature was a result of activity rotating across the solar surface. A number of daily spectra were not considered from the time range due to instrumental or weather effects, the position of these spectra can be seen as solid horizontal lines in the trailed spectrum.

\begin{figure}
		\includegraphics[width=\columnwidth]{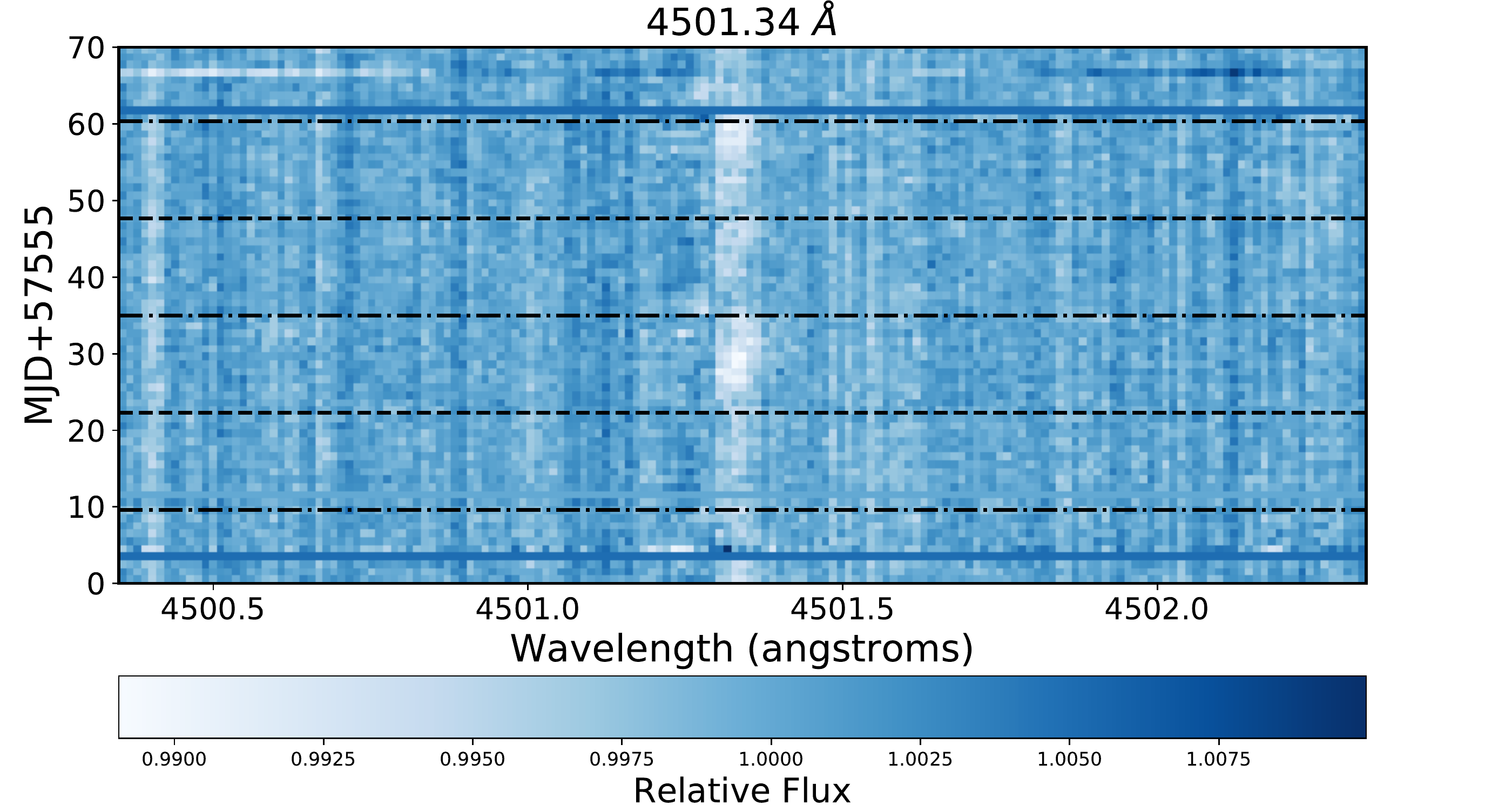}
			\caption{Trailed relative spectrum of the $4501 \angstrom$ line. The difference in colour with respect to Figure~\ref{fig:spec_map4375} is due to this line producing a pseudo-absorption dip. Again, the feature can be seen moving from blueshift to redshift.}
		\label{fig:spec_map4501}
	\end{figure}

Figure~\ref{fig:spec_map4501} shows the same trailed spectrum for the $4501 \angstrom$ feature, which appears as absorption in the relative spectra. Again, this is fully captured within the same $\sim$22 - 35~day half rotation period block and shows the same positive slant -- indicating that the variation in this feature is also driven by changes due to active regions rotating across the solar surface. This RV motion of the pseudo-emission features is similar to that measured for the features in \acb~in Paper~1, further supporting that result. Measurement of these pseudo-emission features then not only give an indication of the level of activity, but also provides velocity information of the active regions themselves, potentially allowing for the active regions to be better localised on the stellar surface. Such velocity information could be highly beneficial in dealing with astrophysical noise in RV measurements, providing a more detailed understanding that could be used to more completely model and remove its effects.

The rest of the features show a similar scenario with the peaks in pseudo-equivalent width  modulated on the solar rotation period, however no indication of RV variation can be seen in any of these other cases.
	
\section{Discussion}\label{sect:discussion}
We show that some of the same `pseudo-emission' features found in \acb~(Paper-1) are also seen in the Sun. This is despite the Sun being in, or approaching, a relatively deep minimum in the solar activity cycle. This work shows that we are sensitive to variations in individual lines in even quiet stars and, from a stellar activity point of view, information gained from studying solar activity can be scaled, at least, to the activity level of \acb. With resolved images of the surface of the Sun, the features can be further characterised as being due to changes in the level of faculae coverage, which is supported by the Pearson's r statistic showing a stronger correlation of the features with faculae than spots -- as well as the stronger correlation to faculae than network using the filling factors from \cite{Milbourne2019}. In Paper~1, we found that assuming spots alone was not enough to produce the distinct pseudo-emission feature morphologies observed in \acb~and that the addition of faculae was needed to more fully recreate the features. The results here further strengthen this conclusion, as we are able to relate the spectral signals detected directly to the appearance of faculae on the solar surface.

In addition, we find that the peak in the pseudo-equivalent widths of the relative features are offset from the spot filling factor. By considering the SDO/HMI images, the peak can be seen to better coincide with the appearance of the extended faculae region and the rotation of this region as it moves from the east to the west limb of the Sun. The second peak in the pseudo-equivalent width can be seen at $\sim$58 days and corresponds to the same extended unassociated faculae region returning. This suggests that regions dominated by faculae, with little to no associated spots (i.e. unassociated facular regions), are the main driver of the features seen in the relative spectra of the Sun.

Although we do not show the magnitude of the RV variation of the pseudo-emission features, the trailed spectra demonstrate that, at least for the two strongest ($4375 \angstrom$ and $4501 \angstrom$), there is a radial velocity component (a result similar to that found in Paper~1). This motion may be a source of astrophysical noise that could impact the RV measurements of planets if not correctly accounted for. The addition of velocity information to these new stellar activity indicators could prove useful in future astrophysical noise correction techniques or as a new set of priors for existing correction techniques that model activity using a Bayesian framework \citep[e.g.][]{Rajpaul2015}.

Other work has shown that faculae dominates inactive stars  \citep{Radick1990, Hall2009, Lockwood2007, Shapiro2016} and  thus will dominate the RV jitter signal \citep{Saar1997, Saar2003, Meunier2010, Dumusque2014, Haywood2016}, the results here for the Sun align with this paradigm. The results also show, however, that the effects of activity are more complex than can be captured by the \logr~of a star. For the Sun at least, faculae that are not associated with spots seem to show a larger impact on the relative spectra and may be the main source of pseudo-emission features seen. From this, one can see that the configurations of active regions (i.e. whether or not spots and faculae are associated) is just as important as the level of coverage. This suggests that the RV jitter due to activity could vary even if the \logr~of the star does not change and could have a significant impact on the use of some stellar jitter reduction techniques. This is most likely to only be applicable for low activity stars where the faculae dominates in activity, but these are also the kind of stars that Earth analogues are most likely to be found around.

While the effects of faculae and plage can dominate the RV jitter of low activity stars, the low contrast of these active regions against the quiet star photosphere means they cause little photometric variability. Methods such as the FF' \citep{Aigrain2012}, for example, accounts for the effects of faculae and plage by assuming that it is related to the spots. The technique assumes that a spot is surrounded by an extended magnetised area that suppresses the convective blue-shift (e.g. plage/faculae) and the RV jitter caused by this region can be estimated by tracking photometic variability due to spots. The results here show that, for low activity stars like the Sun and for certain configurations of activity, the unassociated faculae may generate a more significant impact on the relative spectrum than that of associated spots and faculae.

The nature of the impact of this on methods like the FF' is two-fold. In one instance, if faculae/plage are not associated with spots, then the effects of these regions will not be accounted for by using spots as a proxy for activity. Additionally, as the unassociated faculae is offset from any spots, any RV jitter correction applied that assumes faculae and spots are coincident will have a phase shift from what could be a major contributor to the activity. This could, in turn, inject spurious signals into the RV measurements. Recent work by \cite{Oshagh2017} found that the FF' method systematically under-predicted the amplitude of the RV jitter over the range of activity levels tested, which became significant at low-activity levels, i.e. the level at which plage/faculae effects are expected to dominate. The work here further supports this result, giving a physical interpretation of how activity driven photometric variability may incorrectly account for the RV jitter of low activity stars. 

Photometrically driven methods such as the FF' method are still good first steps in accounting for activity (and still hold for more active stars that are spot dominated). It is possible that the activity indicators presented here could be incorporated into such a physically motivated model to improve the removal process in the low-activity regime.

Although we were able to limit the effects of the strong etaloning seen in the relative spectra, we were not able to fully identify its source. This effect may be resolved, or at the very least better modelled, by performing our own reduction of the raw images that is better optimised for the high signal-to-noise, daily stacked regime we consider. However, this would be a substantial undertaking and outside the scope of this current investigation.

As a brief final note, in Paper~1 we show the existence of two distinct morphologies within the relative features -- those with and without a pseudo-absorption trough surrounding the main peak of the narrow features. In the Sun we did not find conclusive evidence for this ``trough'' morphology. This is most likely due to the lower activity level of the Sun, which produces a dearth of features in the relative spectra when compared to that of \acb. The features that we do see are smaller in strength so more difficult to determine if a trough feature is present. A better model for the etaloning may increase the detail of the features presented here as well as identify other, weaker features. The choice of low-activity template could also be re-visited. Newer data, showing lower levels of activity, could be used instead, which would boost the strength of the features. Both these improvement could not only help determine if the same variation in pseudo-emission feature morphology is seen in the solar case but also identify other lines that may be activity sensitive.

\section{Conclusions}\label{sect:conclusions}

In conclusion, using data from the HARPS-N solar telescope, we produce ``relative spectra'' by dividing high-activity echelle orders by low-activity templates and find, after removing a pattern caused by etaloning, a number of pseudo-emission features. These features correlate well with \logr~and the faculae-filling factors derived from SDO images, but show significantly smaller or no correlation with spot-filling factors or activity indexes published by \cite{Maldonado2019} for the H$\alpha$, H$\beta$, H$\gamma$, H$\delta$, Na {\sc{i}} D1 D2, and He {\sc{i}} D3 lines.

In addition, we focus on a period when a prominent facular region can be seen to rotate onto the visible solar hemisphere at a time when no sunspots are visible. Again, we see that the pseudo-emission features react strongly to the appearance of this facular region -- suggesting that these features are directly tracking changes in faculae coverage. This strengthens the conclusions made in our previous work on \acb~(Paper~1), providing further evidence that the pseudo-emission features track changes in stellar activity and, in particular, faculae. This is important, since it is faculae (and the underlying suppression of the convective blueshift) that is thought to represent the fundamental barrier to RV measurement precisions in the context of exoplanet detection. The fact that the correlation with more standard activity indexes (such as H$\alpha$) is much smaller may be due to a combination of effects -- including telluric line contamination of the H$\alpha$, Na {\sc{i}} D1 D2, and He {\sc{i}} D3 lines; complex line-formation; and that some of these indicators (in particular the Balmer lines) may be more sensitive to filaments.

Finally, we produce trailed spectra and observe the features red-shifting as they move across the solar surface, indicating the pseudo-emission feature have an RV motion. The features are encapsulated within one half of a solar rotation, and further supports our interpretation that the features are driven by activity on the visible surface of the Sun. The pseudo-emission features thus contain both activity and velocity information and could form the basis of an activity indicator that can directly track and remove the effects of faculae to improve planet RV measurements.

\section*{Acknowledgements}
The HARPS-N project has been funded by the Prodex Program of the Swiss Space Office (SSO), the Harvard University Origins of Life Initiative (HUOLI), the Scottish Universities Physics Alliance (SUPA), the University of Geneva, the Smithsonian Astrophysical Observatory (SAO), and the Italian National Astrophysical Institute (INAF), the University of St Andrews, Queen's University Belfast, and the University of Edinburgh.

Based on observations made with the Italian {\it Telescopio Nazionale Galileo} (TNG) operated by the {\it Fundaci\'on Galileo Galilei} (FGG) of the {\it Istituto Nazionale di Astrofisica} (INAF) at the {\it  Observatorio del Roque de los Muchachos} (La Palma, Canary Islands, Spain).

CAW acknowledges support from Science and Technology Facilities Council grant ST/P000312/1.

ACC acknowledges support from the Science and Technology Facilities Council (STFC) consolidated grant number ST/R000824/1.

XD is grateful to the Branco-Weiss Fellowship-Society in Science for its financial support

SHS is grateful for support from NASA Heliophysics LWS grant NNX16AB79G. HMC acknowledges the financial support of the National Centre for Competence in Research PlanetS supported by the Swiss National Science Foundation (SNSF).

AM acknowledges support from the senior Kavli Institute Fellowships.

This work was performed partly under contract with the California Institute of Technology (Caltech)/Jet Propulsion Laboratory (JPL) funded by NASA through the Sagan Fellowship Program executed by the NASA Exoplanet Science Institute (R.D.H.).

This publication was made possible through the support of a grant from the John Templeton Foundation. The opinions expressed in this publication are those of the authors and do not necessarily reflect the views of the John Templeton Foundation. This work was performed in part under contract with the Jet Propulsion Laboratory (JPL) funded by NASA through the Sagan Fellowship Program executed by the NASA Exoplanet Science Institute (in support of R.D.H.).

Finally, we would like to acknowledge the excellent discussions and scientific input by members of the International Team, ``Towards Earth-like Alien Worlds: Know thy star, know thy planet'', supported by the International Space Science Institute (ISSI, Bern).

\bibliographystyle{mnras}
\bibliography{sun_bib} 

% Don't change these lines
\bsp	% typesetting comment
\label{lastpage}

\appendix
\section{Appendix 1}
\begin{figure*}
    \centering
    \includegraphics[width=\textwidth]{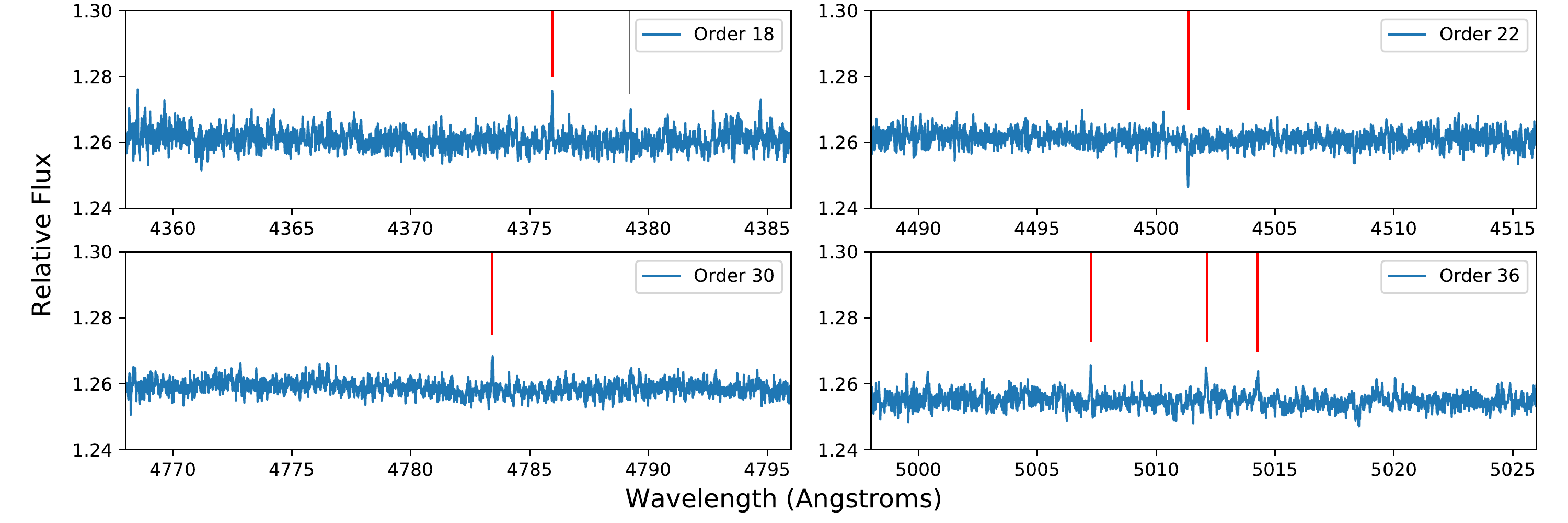}
    \caption{Relative spectra of each of the echelle orders used in this analysis taken from the day with the highest activity as traced by \logr~(that of $14^{\rm{th}}$ July 2016). The vertical lines in red show the positions of identified pseudo-emission (and pseudo-absorption in the case of order 22) features found in the relative spectra that correlate with activity measurements (either \logr~ or spot and faculae filling factors). The exact positions of the features can be found in Table~\ref{table:Sun_lines}. The grey vertical lines shows the position of the 4383$\angstrom$ line, this produces a broad features (which can be seen more clearly in Figure~\ref{fig:Sun_acb_compare}) and not used in this analysis due to difficulties in obtaining an accurate measure of it.}
    \label{fig:orders_plot2}
\end{figure*}
\end{document}